\documentclass[12pt]{article}
\usepackage{epsfig}
\def\th232{\rm{ ^{232} Th }}
\def\u238{\rm{ ^{238} U }}
\def\ur235{\rm{ ^{235} U }}
\def\cs137{\rm{ ^{137} Cs }}

\def\halflife{\tau_{\frac{1}{2}}}

\begin{document}
\hfill AS-TEXONO/05-08\\
\hspace*{1cm} \hfill \today

\begin{center}
{\Large  
\bf{
Measurement of the Intrinsic Radiopurity \\
of $^{137}$Cs/$^{235}$U/$^{238}$U/$^{232}$Th in \\
CsI(Tl) Crystal Scintillators
}\\
}
\vskip 0.2cm
\large
Y.F.~Zhu$^{a,b}$,
S.T.~Lin$^{b,c}$,
V.~Singh$^{b,d}$,
W.C.~Chang$^{b}$,
M.~Deniz$^{b,e}$,\\
W.P.~Lai$^{b}$,
H.B.~Li$^{b}$,
J.~Li$^{a,b,f}$,
Y.L.~Li$^{a}$,
H.Y. Liao$^{b,c}$,\\
F.S.~Lee$^{b}$,
H.T.~Wong$^{b,}$\footnote{Corresponding~author $-$\\
\hspace*{1cm} Mailing Address: 
Institute of Physics, Academia Sinica, Taipei 11529, Taiwan;\\
\hspace*{1cm} Email:~htwong@phys.sinica.edu.tw;
Tel:+886-2-2789-6789;
FAX:+886-2-2788-9828.},
S.C.~Wu$^{b}$,
Q.~Yue$^{a}$
Z.Y.~Zhou$^{g}$\\
\vskip .2cm
The TEXONO Collaboration
\end{center}

\normalsize

\begin{flushleft}
{$^{a}$\rm 
Department of Engineering Physics, Tsing Hua University, 
Beijing.\\}
{$^{b}$\rm 
Institute of Physics, Academia Sinica, Taipei.\\}
{$^{c}$\rm
Department of Physics, National Taiwan University, Taipei.\\}
{$^{d}$\rm
Department of Physics, Banaras Hindu University, Varanasi.\\}
{$^{e}$\rm
Department of Physics, Middle East Technical University, Ankara.\\}
{$^{f}$\rm 
Institute of High Energy Physics, Chinese Academy of Science,
Beijing.\\}
{$^{g}$\rm
Department of Nuclear Physics,
Institute of Atomic Energy, Beijing.\\}
\end{flushleft}

\vfill
\newpage

\begin{abstract}

The inorganic crystal scintillator CsI(Tl) 
has been used for low energy neutrino and
Dark Matter experiments, where the
intrinsic radiopurity is an issue of major 
importance.
Low-background data were taken with
a CsI(Tl) crystal array 
at the Kuo-Sheng Reactor Neutrino Laboratory.
The pulse shape discrimination capabilities of the
crystal, as well as 
the temporal and spatial correlations of the events,
provide powerful means of measuring 
the intrinsic radiopurity of  
$\cs137$ as well as the  
$\ur235$, $\u238$ and $\th232$ series.
The event selection algorithms are described,
with which the 
decay half-lives of $^{218}$Po,
$^{214}$Po, $^{220}$Rn, $^{216}$Po
and $^{212}$Po were derived.
The measurements of the 
contamination levels, 
their concentration gradients with
the crystal growth axis, and the uniformity among
different crystal samples, are reported. 
The radiopurity
in the $\u238$ and $\th232$ series 
are comparable to those of the best reported in 
other crystal scintillators.
Significant improvements 
in measurement sensitivities were achieved,
similar to those from
dedicated massive liquid scintillator detector.
This analysis also provides in situ
measurements of the detector performance
parameters, such
as spatial resolution, quenching factors,
and data acquisition dead time.

\end{abstract}


\begin{flushleft}
{\bf PACS Codes:}  
29.40.Mc, 29.25.R, 07.05.Kf.\\
{\bf Keywords:}
Scintillation detector, Radioactive sources, Data analysis.
\end{flushleft}

\vfill

\newpage

\normalsize

\section{Introduction}

The merits and physics potentials
for crystal scintillators in low-energy
low-background experiments were recently
discussed~\cite{prospects}.
This detector technology has been
adopted in double beta decay, dark matter
and reactor neutrino experiments, and
considered for low energy solar neutrino
detection.
An important common issue is the
contamination of radioactive
isotopes in the crystals.
This article reports measurements of
the intrinsic
radiopurity in CsI(Tl) crystals.

The measurements were performed
at the Kuo-Sheng (KS) Nuclear Power Plant
where a laboratory 
located at a distance of 28~m from the 
reactor core has been built
to study low energy
neutrino physics~\cite{ksprogram}.
A multi-purpose ``inner target'' detector space of
100~cm$\times$80~cm$\times$75~cm is
enclosed by 4$\pi$ passive shielding materials
which have a total weight of 50 tons.
The shielding provides attenuation
to the ambient neutron and gamma background, and
consists of, from inside out,
5~cm of OFHC copper, 25~cm of boron-loaded
polyethylene, 5~cm of steel, 15~cm of lead,
and cosmic-ray veto scintillator panels.

Different detectors can be placed in the
inner space for different scientific goals.
The detectors are read out by a versatile
electronics and data acquisition (DAQ) system\cite{eledaq}
based on 16-channel, 20~MHz, 8-bit
Flash Analog-to-Digital-Convertor~(FADC) modules.
The readout allows full recording of all the relevant pulse
shape and timing information for as long as 500~$\mu$s
after the initial trigger
with a resolution of 1~$\mu$s.
The data are then read out and transferred to the
host computer, and the average DAQ dead time is
5.13~ms per event. 
The timing information for every event
was recorded with a resolution of $\sim$1~ms.
The time differences between
subsequent events and between different
triggers within an event can be evaluated.
Software procedures have been devised to extend the
effective dynamic range from the  nominal 8-bit
measurement range provided by the FADC\cite{dyrange}.

\begin{figure}
\centerline{
\epsfig{file=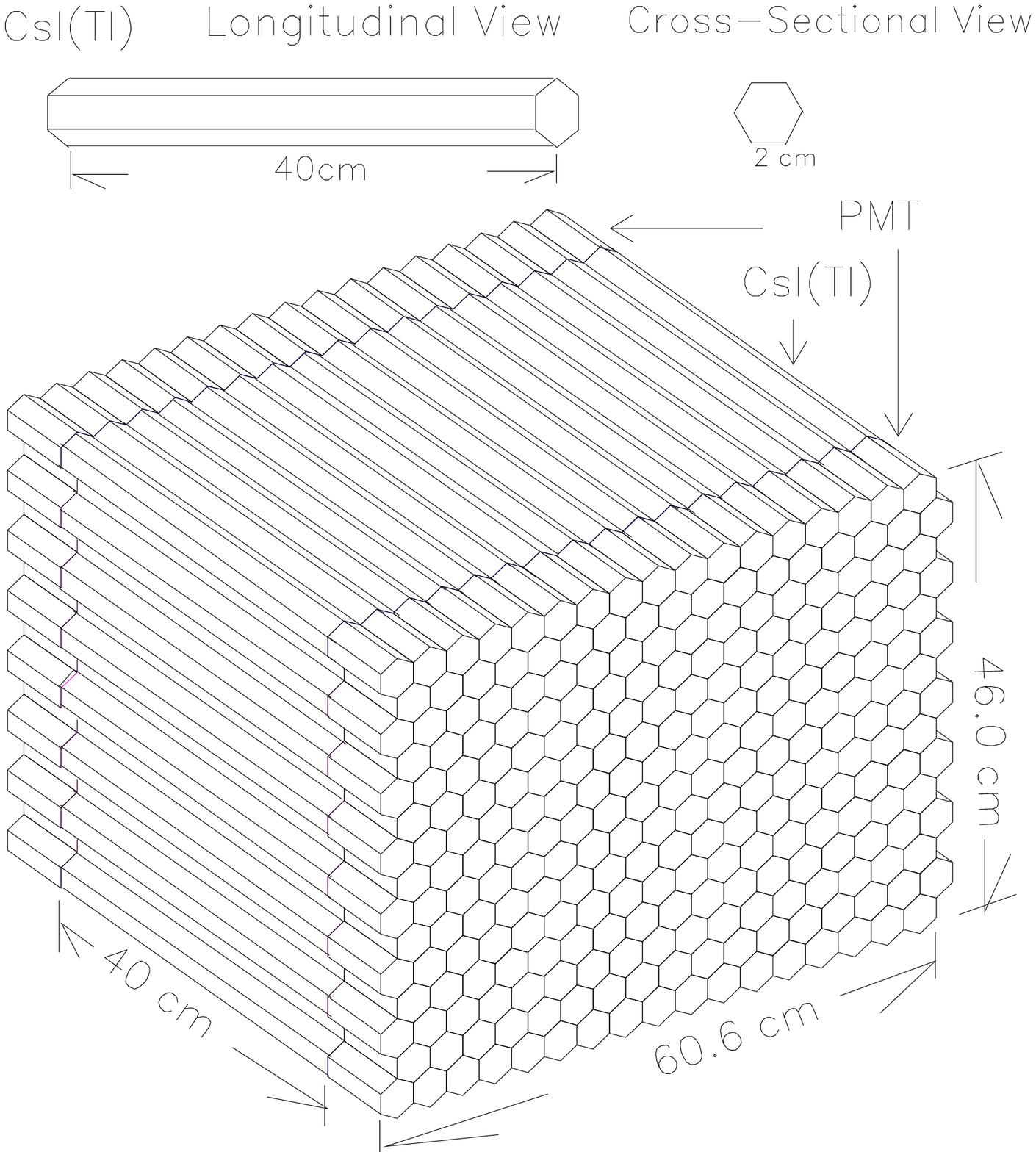,width=14cm}
}
\caption{
Schematics of 
the CsI(Tl) target configuration where
a total of 93 modules (186~kg)
is installed for the 2003-04 data taking.
}
\label{csiarray}
\end{figure}

One of the detectors placed inside the inner
target volume is an array of CsI(Tl) crystal
scintillators\cite{kscsi}, 
the schematic of which is
displayed in Figure~\ref{csiarray}.
Each crystal module
is 2~kg in mass and
consists of a hexagonal-shaped cross-section with 2~cm
side and a length of 40~cm.
The light output are read out
at both ends ($\rm{Q_R ~ and ~ Q_L}$) 
by custom-designed 29~mm diameter
photo-multiplier tubes(PMTs) with low-activity glass.
The sum and difference of the PMT signals give information
on the energy and the longitudinal Z-position of
the events, respectively. Performance of the
prototype modules was published in
Ref.~\cite{proto}.

A total of 186~kg (or 93~modules) was
commissioned for the 2003 data taking.
The physics goal is to measure
the Standard Model neutrino-electron
scattering cross sections, and thereby
to provide a measurement of $\rm{sin ^2 \theta _W}$
at the untested MeV range.
The strategy~\cite{sensit} is
to focus on data at high ($>$2~MeV) recoil
energy where the uncertainties due to
the reactor neutrino spectra are small.
The large target mass compensates the
drop in the cross-sections at high energy.

This article focuses on the measurements of the
intrinsic $\cs137$ level as well as
those for the $\ur235$,
$\u238$ and $\th232$ series
in the CsI(Tl) crystals using the correlated events 
from $\beta$-$\alpha$ and $\alpha$-$\alpha$ decay
sequences.
Our work complements and extends the scope of those
previous efforts on this~\cite{csibkg,kims}
and other crystal scintillators:
NaI(Tl)~\cite{naidama,naiuk,naijapan}, 
CdWO$_4$~\cite{cdwo4}, GSO(Ce)~\cite{gso},
CeF$_3$~\cite{cef3},
BaF$_2$~\cite{baf2} and CaF$_2$(Eu)~\cite{caf2}.
The achieved sensitivities 
match the best among these results
as well as those from massive 
liquid scintillator facility~\cite{ctf}.
Several detector parameters
can be derived from these measurements,
providing a means of {\it in situ} monitoring of
the detector calibration and performance.

In addition, another CsI(Tl) detector
array of mass 200~kg is being prepared
in the Yang-Yang Underground Laboratory
in South Korea to look for Cold Dark Matter~\cite{kims}.
Knowledge of the intrinsic radioactivity
in CsI(Tl) is essential to both
experiments. The methods devised are applicable
to a wide spectrum of low-background experiments
where $\alpha$/$\gamma$ discrimination is possible
and the detailed timing information are available.

\section{Intrinsic  $^{137}$Cs Level} 

The isotope $\cs137$ is produced artificially
as fission waste from power reactors and atomic
weapon tests. Cesium compounds are highly soluble
and hence $\cs137$ can easily be introduced
into cesium products during the processing from
cesium ore (like pollucite). Once introduced,
there is no chemical means to separate 
$\cs137$ from the stable $^{133}$Cs.
This is the dominant
one background contribution
to Dark Matter experiments
using CsI(Tl) crystals as target~\cite{kims}.

\begin{figure}
\centerline{
\epsfig{file=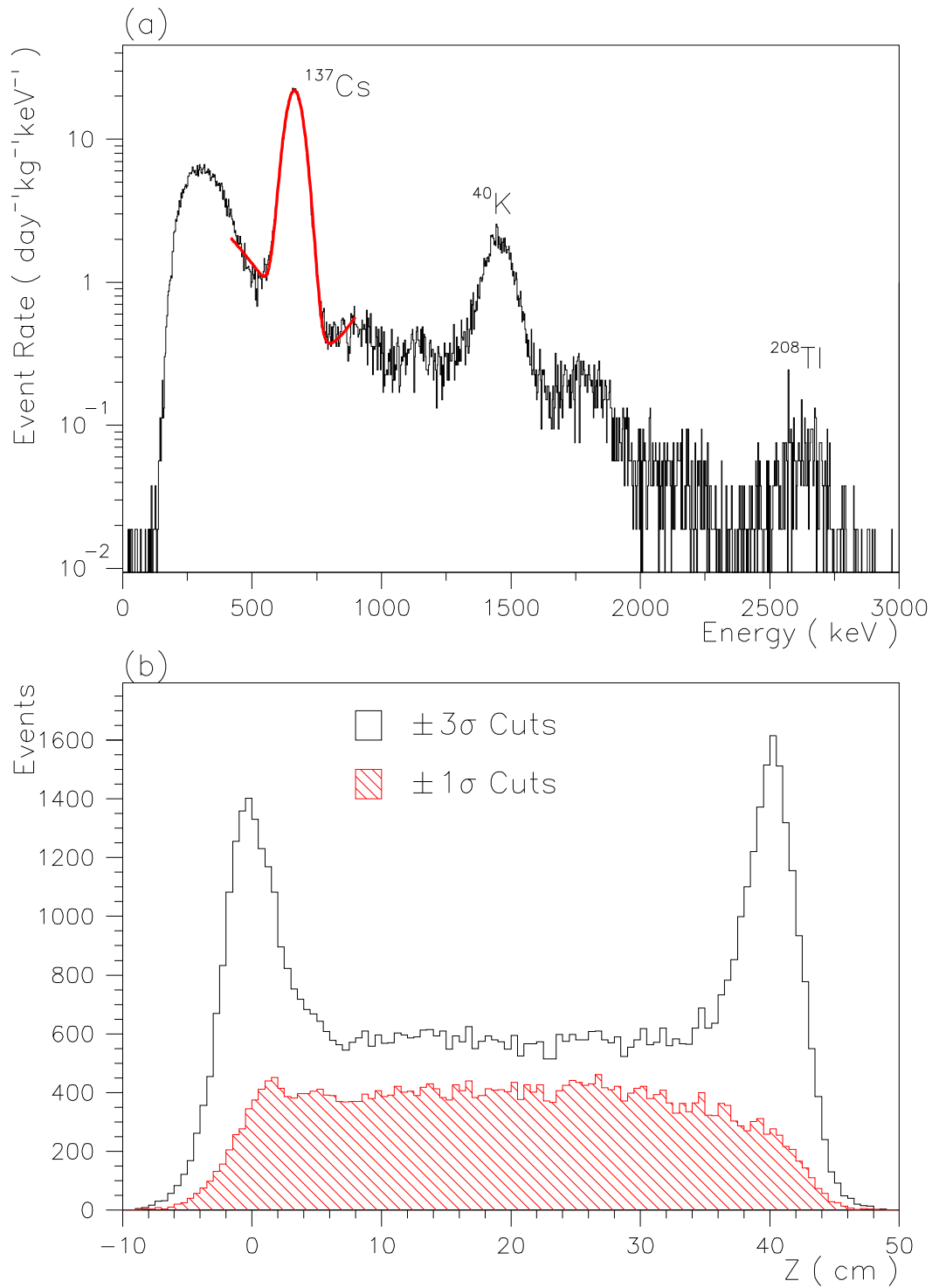,width=14cm}
}
\caption{
(a) Energy spectrum of the entire range 
and (b) reconstructed Z-position
distribution of the $^{137}$Cs events
with 31.3~kg-day of data.
}
\label{cs137}
\end{figure}

The isotope decays via~\cite{tori}:
\begin{eqnarray*}
^{137}Cs ~ & \rightarrow & ~ ^{137}Ba^* ~ + ~ e^- ~ 
( \halflife =30.07~y ) ~ ,\\
^{137}Ba^* ~ & \rightarrow & ~ ^{137}Ba ~ + ~ \gamma ~
( E_{\gamma} = 662~keV ) ~ 
\end{eqnarray*}
with the emission of a $\gamma$-ray of energy 
662~keV.
A total of 31.3~kg-day of CsI(Tl) data was analyzed.
The spectrum is displayed in Figure~\ref{cs137}a,
showing a distinct line at this energy
with a root-mean-square (RMS) resolution 
of 5.1\%. Other naturally-occuring $\gamma$-lines,
like those from $^{40}$K and $^{208}$Tl, 
can also be observed.

The Z-position
distribution of the $^{137}$Cs events
are shown in Figure~\ref{cs137}b.
Events within $\pm 3 \sigma$ of the
$\gamma$-peak give excess 
on both ends of the 40-cm long crystals due to
external background.
The uniform distribution after a $\pm 1 \sigma$
cut indicates the source of $\cs137$ is
internal to the crystals. 

Simulation studies give an efficiency of 42\%
for full capture of the 662~keV $\gamma$-ray
in a single crystal. Accordingly, an average
activity of 61$\pm$2~mBq kg$^{-1}$, or equivalently
a contamination level of 
$( 1.55  \pm 0.05 ) \times 10^{-17}$~g/g, 
can be derived
for the CsI(Tl) crystals. 
The RMS spread of the 
activity among the 31~crystals studied
is 30\%.

Alternatively, the radiopurity of the
CsI powder with which the CsI(Tl) crystals
were produced have been measured with 
a high-purity germanium detector.
The measured $\cs137$ abundance 
is $( 1.7 \pm 0.3 ) \times 10^{-17}$~g/g, consistent
with the in situ results.

\section{Pulse Shape Discrimination in CsI(Tl) Crystals}
\label{sect::psd}

\begin{figure}
\centerline{
\epsfig{file=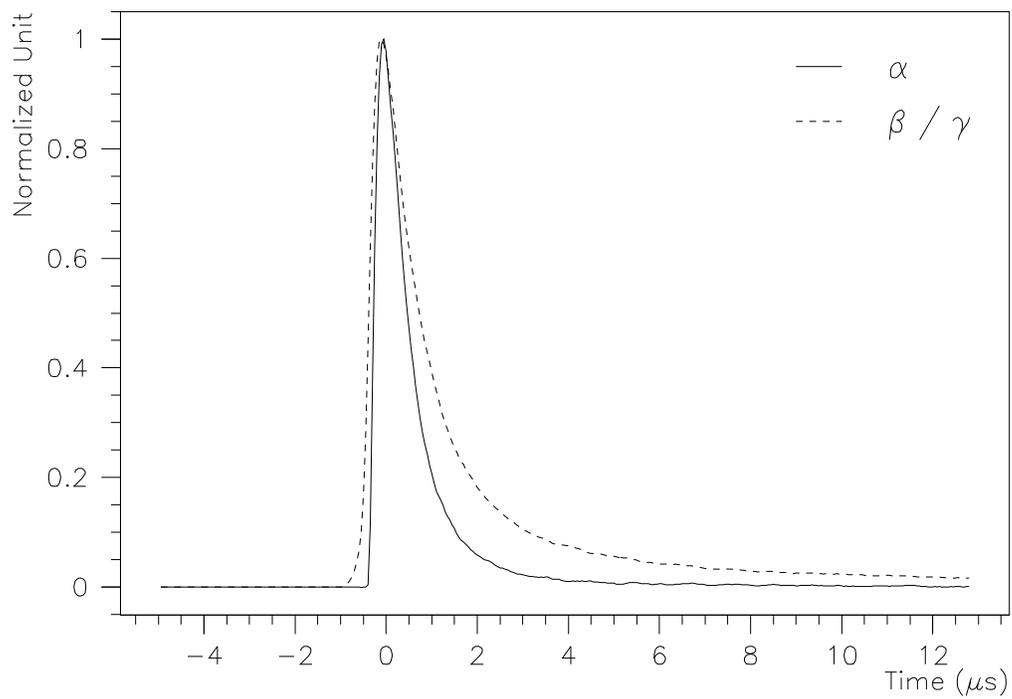,width=14cm} 
}
\caption{
Average pulse shapes for
$\alpha$ and $\beta$/$\gamma$ events 
}
\label{psdshape}
\end{figure}

It has been well-studied~\cite{proto,csichar}
that the light emission profiles of
scintillating CsI(Tl) crystals
exhibit different shape for
$\gamma$-rays and electrons (that is, minimum ionizing particles),
as compared to that for
$\alpha$-particles and nuclear recoils at
the $>$100~keV regime~.
Heavily ionizing
events  due to $\alpha$-particles and nuclear
recoils have
{\it faster} decays than those from e/$\gamma$'s $-$
opposite to the response in
liquid scintillator~\cite{scinbasic}.
The {\it average} pulse shapes for both categories
are depicted in Figures~\ref{psdshape},
where t=0 is defined by the trigger instant.
This characteristic property makes particle identification
possible with this scintillator~\cite{psdpid}.

Matured pulse shape discrimination (PSD)
techniques have been devised
at high energies where the photo-electrons
are abundant.
The ``double charge method''~\cite{psddc}
involves the comparison of
the ``total charge'' ($\rm{Q_{tot}}$)
and the ``partial charge'' ($\rm{Q_{part}}$),
which are the total and partial
integration of the pulse, respectively.
In this analysis, $\rm{Q_{tot}}$ and
$\rm{Q_{part}}$ were evaluated
by integrating the entire pulse 
from $-$0.75 to 12.8~$\mu$s,
and the tail from 5 to 12.8~$\mu$s, respectively, with respect to
the timing shown in
Figure~\ref{psdshape}.

\begin{figure}
\centerline{
\epsfig{file=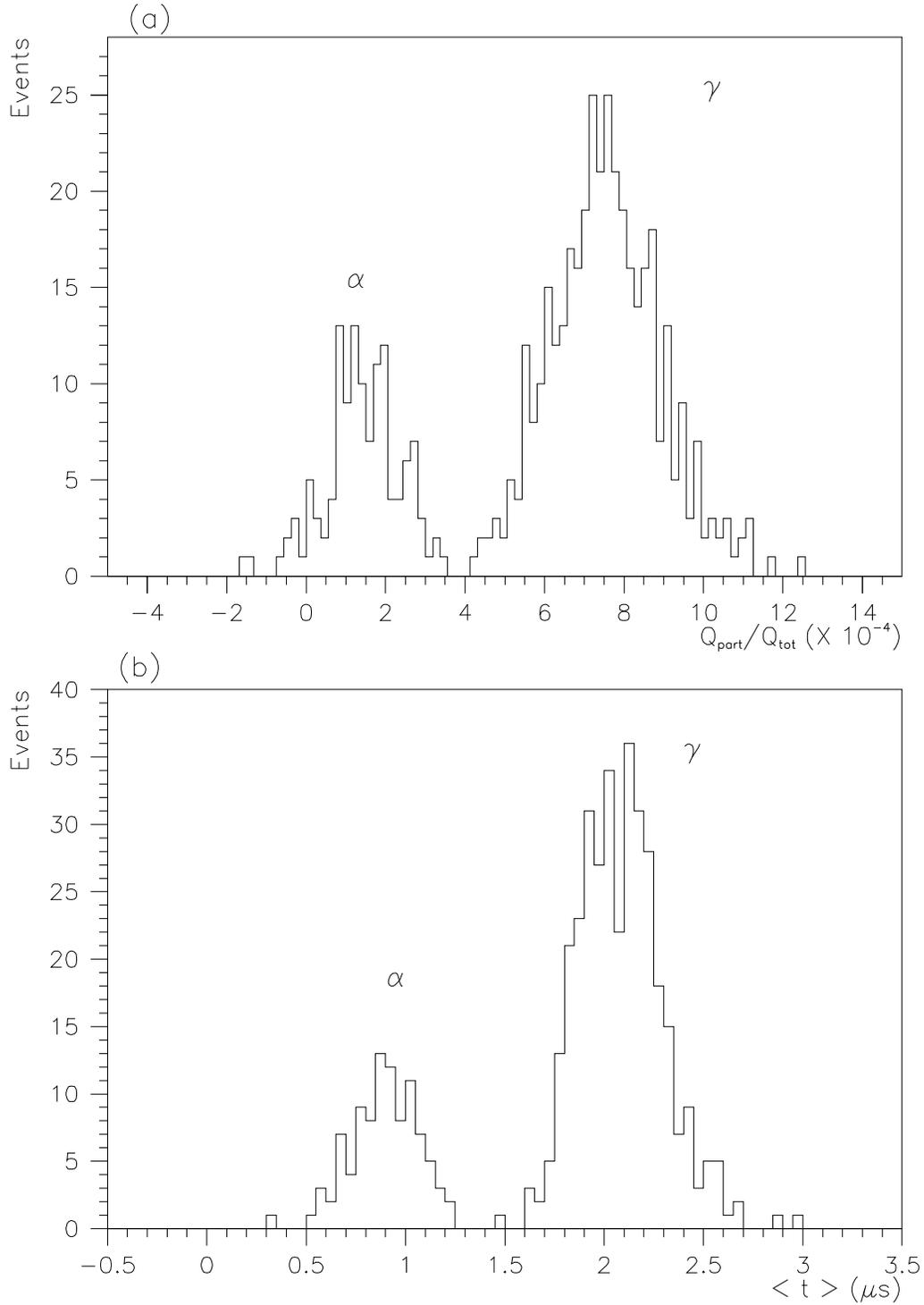,width=14cm}
}
\caption{
Event identification with
Pulse Shape Discrimination with:
(a)
the partial charge versus total charge method,
and
(b)
the mean time method.
The $\gamma$ samples
are from $^{40}$K, while the
$\alpha$ events are with 
electron-equivalence energy of more
than 1~MeV. A perfect event identification
is thus possible.
}
\label{psd}
\end{figure}

With the digitized pulse information
measured by the FADCs, the
mean time method is often used.
The mean time is defined as
\begin{equation}
\rm{
\langle t \rangle ~ = ~
 { \sum\limits_{i} ~ ( A_i ~  t_i ) \over \sum\limits_{i} ~ A_i }  ~~ ,
}
\end{equation}
where $\rm{A _i}$ is the FADC-amplitude at
time-bin $\rm{t _i}$.
The PSD capabilities to make $\alpha$/$\gamma$
identification for MeV events are displayed
in Figure~\ref{psd}a\&b for both methods,
indicating excellent separation.
These features can be extended to the
low energy regime 
where photo-electron statistics
are limited, via
advanced software techniques~\cite{lepsd}.
This is relevant to the Dark Matter experiments.

\section{Intrinsic Radiopurity of U/Th Series}

The $\ur235$, $\u238$ and $\th232$ decay series 
are naturally occurring, and exist in all materials
at some levels. The processing of the raw
materials to CsI(Tl) crystals can preferentially
introduce or remove certain isotopes within
these series such that secular equilibrium
can be destroyed.
A sensitive measurement of  
the spatially and 
temporally-correlated events within
the decay series in CsI(Tl) crystals
was performed. 
The measured activities were translated to
contamination levels of their long-lived
parent isotopes  in the crystal.
If secular equilibrium is assumed,
the levels of 
$\ur235$, $\u238$ and $\th232$ 
can be derived.

The identification of the
events provides event-by-event background
rejection to the candidate events for
neutrino or Dark Matter-induced interactions.
Beyond that, 
they are extremely useful to perform
{\it in situ}
studies of detector performance parameters.

\subsection{Decay Sequences}
\label{sect::cascade}

The five decay sequences (DS) of
interest under studied are~\cite{tori}:
\begin{description}
\item{\bf DS$_1$)~} Within the $\u238$ series,
the $\alpha$-$\alpha$ cascade from $^{222}$Rn $-$
\begin{eqnarray*}
^{222}Rn ~ & \rightarrow & ~ ^{218}Po ~ + ~ \alpha ~
(Q=5.59~MeV ~ ; ~ \halflife = 3.82~d ) ~ , \\
^{218}Po ~ & \rightarrow & ~ ^{214}Pb~ + ~ \alpha ~
(Q=6.12~MeV ~ ; ~ \halflife =3.10~min ) ~~~,
\end{eqnarray*}
the event rate of which gives the concentration of
$^{226}$Ra whose half-life ($\halflife$) is 1600~y.

\item{\bf DS$_2$)~} Within the $\u238$ series, 
the $\beta$-$\alpha$ cascade from $^{214}$Bi $-$
\begin{eqnarray*}
^{214}Bi ~ & \rightarrow & ~ ^{214}Po ~ + ~ \bar{\nu_e} ~ +
~ e^- ~ + ~ \gamma 's ~  (Q=3.28~MeV ~ ; ~ \halflife = 19.8~min) ~, \\
^{214}Po ~ & \rightarrow & ~ ^{210}Pb~ + ~ \alpha ~
(Q=7.83~MeV ~ ; ~ \halflife =164~ \mu s) ~~~,
\end{eqnarray*}
the event rate of which gives the concentration of
$^{226}$Ra with $\halflife$=1600~y.

\item{\bf DS$_{3a,3b}$)~} The decay sequence
$^{224}$Ra$\rightarrow$$^{220}$Rn$\rightarrow$$^{216}$Po
within the $\th232$ series
can be identified by a triple
$\alpha$-$\alpha$-$\alpha$ signature $-$
\begin{eqnarray*}
^{224}Ra ~ & \rightarrow & ~ ^{220}Rn~ + ~ \alpha ~
(Q=5.79~MeV ~ ; ~ \halflife =3.66~d) ~, \\
^{220}Rn ~ & \rightarrow & ~ ^{216}Po~ + ~ \alpha ~
(Q=6.41~MeV ~ ; ~ \halflife =55~s) ~, \\
^{216}Po ~ & \rightarrow & ~ ^{212}Pb~ + ~ \alpha ~
(Q=6.91~MeV ~ ; ~  \halflife  =0.15~s) ~,
\end{eqnarray*}
the event rate of which gives the concentration of
$^{228}$Th with $\halflife$=1.9~y.

\item{\bf DS$_4$)~} Within the $\th232$ series,
there is 64\% branching ratio for $^{212}$Bi
to decay via a $\beta$-$\alpha$ cascade $-$
\begin{eqnarray*}
^{212}Bi ~ & \rightarrow & ~ ^{212}Po ~ + ~ \bar{\nu_e} ~ +
~ e^- ~ + ~ \gamma 's ~  (Q=2.25~MeV ~ ; ~ 
\halflife =60.6~min) ~, \\
^{212}Po ~ & \rightarrow & ~ ^{208}Pb~ + ~ \alpha ~
(Q=8.95~MeV ~ ; ~ \halflife =0.30~ \mu s) ~~~,
\end{eqnarray*}
the event rate of which also gives the concentration of
$^{228}$Th.

\item{\bf DS$_{5a,5b}$)~} Though not expected to be dominant,
the decay chain of
$^{223}$Ra$\rightarrow$$^{219}$Rn$\rightarrow$$^{215}$Po
in the $^{235}$U series has 
a triple $\alpha$-$\alpha$-$\alpha$ sequence $-$
\begin{eqnarray*}
^{223}Ra ~ & \rightarrow & ~ ^{219}Rn ~ + ~ \alpha ~
(Q=5.97~MeV ~ ; ~ \halflife =11.4~d) ~,\\
^{219}Rn ~ & \rightarrow & ~ ^{215}Po~ + ~ \alpha ~
(Q=6.94~MeV ~ ; ~ \halflife =3.96~s) ~,\\
^{215}Po ~ & \rightarrow & ~ ^{211}Pb~ + ~ \alpha ~
(Q=7.52~MeV ~ ; ~ \halflife =1.78~ms) ~~~,
\end{eqnarray*}
the event rate of which  gives the concentration of
$^{227}$Th with $\halflife$=21.8~y.

\end{description}

\subsection{Event Selection}
\label{sect::eselect}

A total of 40~crystals with a data size
of 1725~kg-day taken at the KS Lab was studied.
Various selection criteria 
for the decay sequences DS$_{1-5}$  were devised
and their efficiencies were measured, as summarized
in Table~\ref{dsselect}.
With the exception of the fast-cascade in DS$_{4}$,
the events were identified as $\alpha$ or e/$\gamma$ (ID);
the delayed events should be within four half-lives
from the precedent ones ($\Delta$t); the measured
energy of the $\alpha$-events should match that of
expectations from Quenching Factor
measurements (E$_{\alpha}$) while the
correlated events should have the same Z ($\Delta$Z=0).
Events from DS$_4$ were identified by the
convoluted $\beta$-$\alpha$ pulse shapes.

\begin{table}[hbt]
\begin{center}
\begin{tabular}{lclcccc}
\hline
DS & Signatures & Decays & Selection & 
Background &
\multicolumn{2}{c}{Half-Life ($\tau_{\frac{1}{2}}$)} \\ 
& & & Efficiency & -to-Signal & Nominal & Measured \\\hline \hline
1 & $\alpha$-$\alpha$ &
$\rm{^{222}Rn \rightarrow ^{218}Po}$ & & & &  \\
& & $\rm{~~~~~~~ \rightarrow ^{214}Pb}$ &
0.93 & 0.51  & 3.10~min & 3.2$\pm$0.2~min  \\ \hline
2 & $\beta$-$\alpha$ &
$\rm{^{214}Bi \rightarrow ^{214}Po }$ & & & & \\
& & $\rm{~~~~~~~ \rightarrow ^{210}Pb}$ &
0.77 & $3.2 \times 10^{-3}$ & 164~$\mu$s & 163$\pm$8~$\mu$s \\ \hline
3a & $\alpha$-$\alpha$ &
$\rm{^{224}Ra \rightarrow ^{220}Rn}$ & & & & \\
& & $\rm{~~~~~~~ \rightarrow ^{216}Po}$ &
0.86 & & 55~s &  54.4$\pm$2.4~s  \\
3b & $\alpha$ &
$\rm{~~~~~~~ \rightarrow ^{212}Pb}$ &
0.97 & $9 \times 10^{-5}$ & 0.15~s & 0.141$\pm$0.006~s \\ \hline
4 & $\beta$-$\alpha$ &
$\rm{^{212}Bi \rightarrow ^{212}Po }$ & & & & \\
& & $\rm{~~~~~~~ \rightarrow ^{208}Pb}$ &
0.37 & $3 \times 10^{-5}$ & 299~ns & 283$\pm$37~ns \\ \hline
5a & $\alpha$-$\alpha$ &
$\rm{^{223}Ra \rightarrow ^{219}Rn }$ & & & & \\
& & $\rm{~~~~~~~ \rightarrow ^{215}Po}$ &
0.78 & $-$  & 3.96~s &  No Signal \\ 
5b & $\alpha$ & $\rm{~~~~~~~ \rightarrow ^{211}Pb}$ & 
$-$ & $-$ & 1.78~ms & DAQ Inactive \\ \hline
\end{tabular}
\caption{
\label{dsselect}
Summary of the selection efficiencies, background levels 
and measured
half-lives of the decay sequences.}
\end{center}
\end{table}

\begin{figure}
\centerline{
\epsfig{file=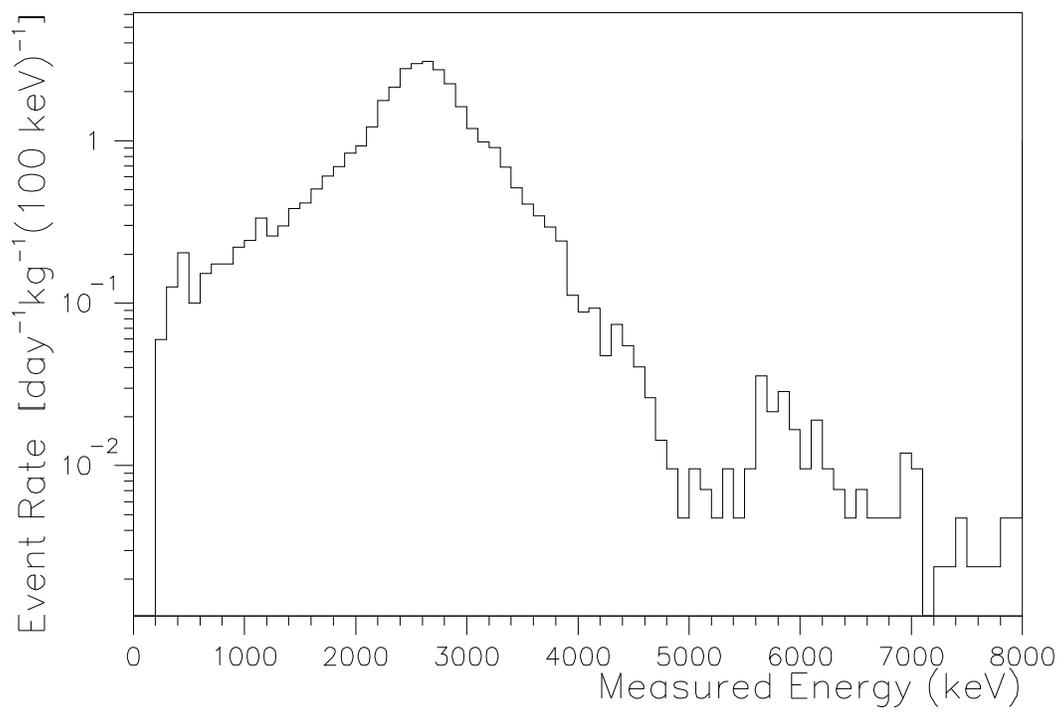,width=14cm}
}
\caption{
Measured energy spectrum for
the $\alpha$-events.
}
\label{alphaspectrum}
\end{figure}

The energy spectrum of the
$\alpha$-events is depicted in
Figure~\ref{alphaspectrum}.
The measured energy is ``quenched'' 
due to
recombinations between the ions
and electrons such that  it
is less than that of $\gamma$/electrons
at the same kinetic energy. The
quenching factor measurements are 
discussed in details in Section~\ref{subsec::qf}.
The small peak at 6~MeV measured energy is due
to the energy sum of those
$\beta$-$\alpha$ pairs in DS$_4$
whose $\Delta$t are too small to
be de-convoluted. 
Apart from this structure, 
the $\alpha$-spectrum shows a broad peak
at about 2.5~MeV.
The individual mono-energetic $\alpha$-emissions
within the $\u238$ and $\th232$ series cannot 
be resolved. This poses difficulties
and limits the accuracies
to radiopurity measurements based
only on the $\alpha$-spectra
without the timing information.

The single $\alpha$ rate 
is 0.0007~kg$^{-1}$s$^{-1}$
or one event per crystal per 11.9~min.
This defines a time-range under which the
identification of correlated $\alpha$-pairs 
would be possible, and justifies
our selection of the five decay sequences
for investigations. 

\begin{figure}
\centerline{
\epsfig{figure=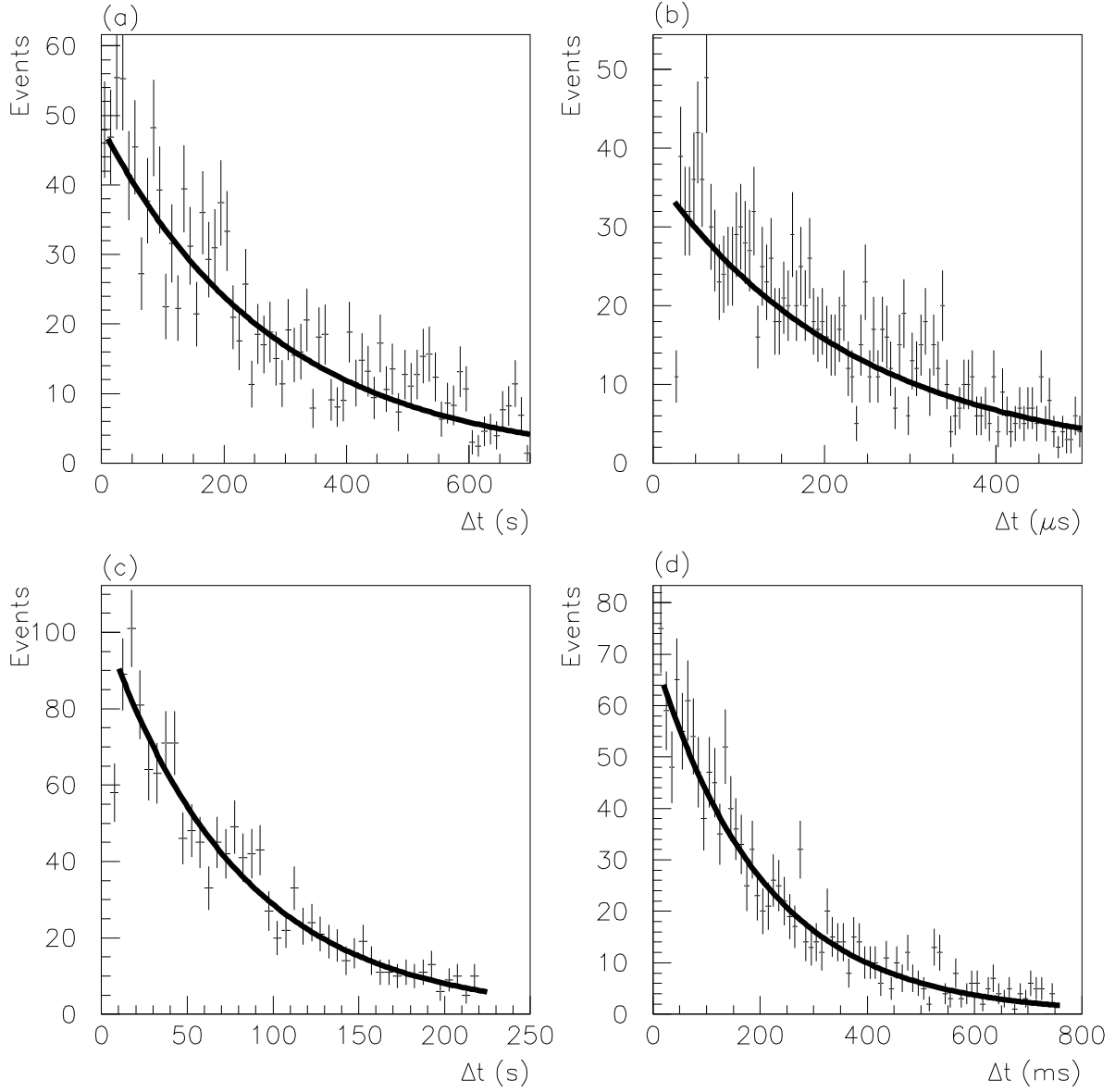,width=16cm}
}
\caption{Delay time ($\Delta$t) distribution for
(a) $\alpha$-$\alpha$ events from
$^{222}$Rn$\rightarrow$$^{218}$Po$\rightarrow$$^{214}$Pb
in DS$_{1}$; 
(b) $\beta$-$\alpha$  events from
$^{214}$Bi$\rightarrow$$^{214}$Po$\rightarrow$$^{214}$Po 
in DS$_2$;
(c) $\alpha$-$\alpha$ events from
$^{224}$Ra$\rightarrow$$^{220}$Rn$\rightarrow$$^{216}$Po
in DS$_{3a}$; 
(d) $\alpha$-$\alpha$ events from
$^{220}$Rn$\rightarrow$$^{216}$Po$\rightarrow$$^{212}$Pb
in DS$_{3b}$. 
}
\label{delaytime}
\end{figure}

The $\Delta$t distributions for
the correlated-events from DS$_{1,2,3a,3b}$
are shown in Figure~\ref{delaytime}a-d, respectively.
Events with delay up to four 
nominal decay half-lives 
were sampled, except
for the DS$_2$ $\beta$-$\alpha$ pairs 
in Figure~\ref{delaytime}b,
where the maximum delay between triggers is 
500~$\mu$s constrained by the DAQ settings. 
In Figure~\ref{delaytime}a where
DS$_1$ $\alpha$-pairs 
with delay time as long as 
12~minutes  were selected,
background due to accidental
pairs and those from DS$_{3a}$ with missing DS$_{3b}$ 
were statistically subtracted. 
Selections for the
other sequences followed from
the criteria discussed. 
The $\Delta$t distributions are compared to
exponential decay profiles, and 
the best-fit half-lives are tabulated
in Table~\ref{dsselect}.
All measurements are in 
excellent agreement with the standard values.

\begin{figure}
{\bf (a)}\\
\centerline{
\epsfig{figure=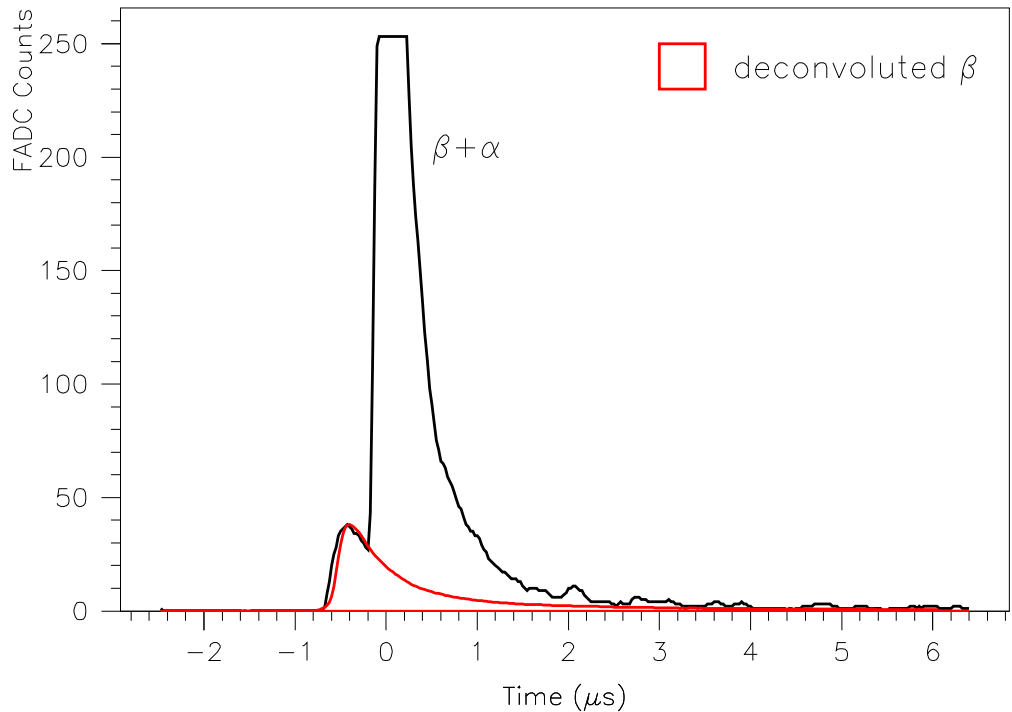,width=14cm}
}
{\bf (b)}\\
\centerline{
\epsfig{file=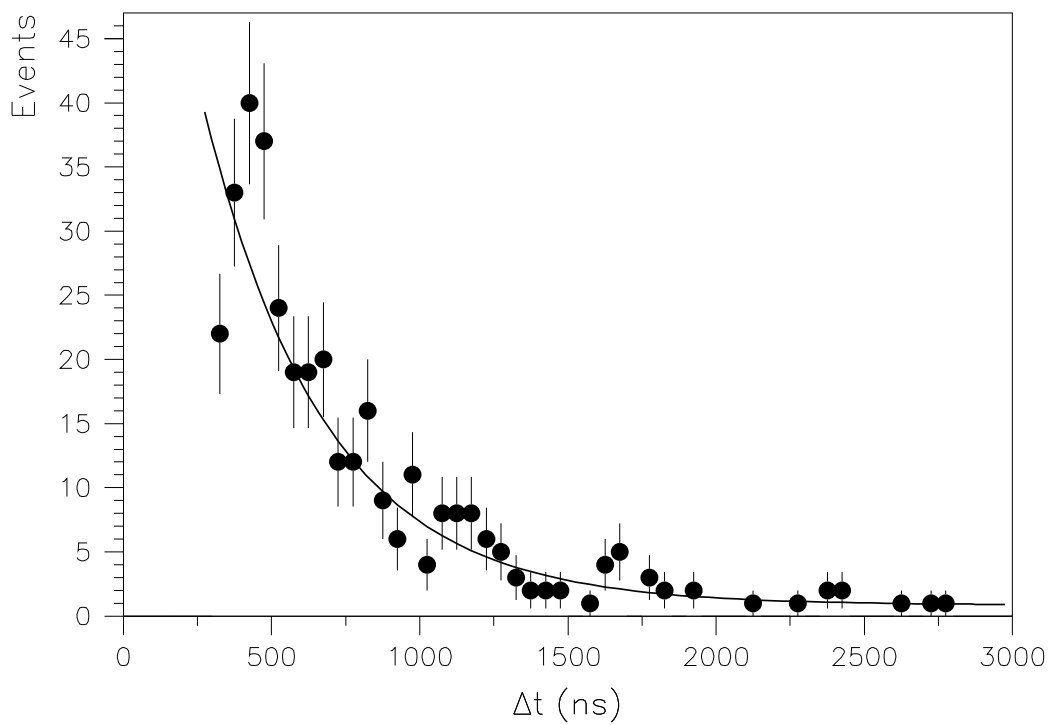,width=14cm}
}
\caption{
(a)
A typical $\beta$-$\alpha$
signal of the cascade
$^{212}$Bi$\rightarrow$$^{212}$Po$\rightarrow$$^{208}$Pb
(b) 
The measured time difference between
the leading edges in the
$\beta$ and $\alpha$ pulses.
}
\label{ds4}
\end{figure}

The sequence DS$_4$ has $\halflife$ of only
300~ns, such that both pulses will appear
convoluted within the same readout time
window of 12.5~$\mu$s. An example of such
signatures is displayed in Figure~\ref{ds4}a.
The cascade was therefore selected not by
timing but by pulse shape analysis.
The time difference between the $\beta$
and $\alpha$ leading edges is depicted
in Figure~\ref{ds4}b, where the best-fit
$\halflife$ is 283$\pm$37~ns, also in
good agreement with reference values. 
As shown in Table~\ref{dsselect}, only 37\%
of the $\beta$-$\alpha$ events can be resolved.
The remaining ones are convoluted and
contribute to the 6~MeV peak in Figure~\ref{alphaspectrum}.

\begin{figure}
\centerline{
\epsfig{figure=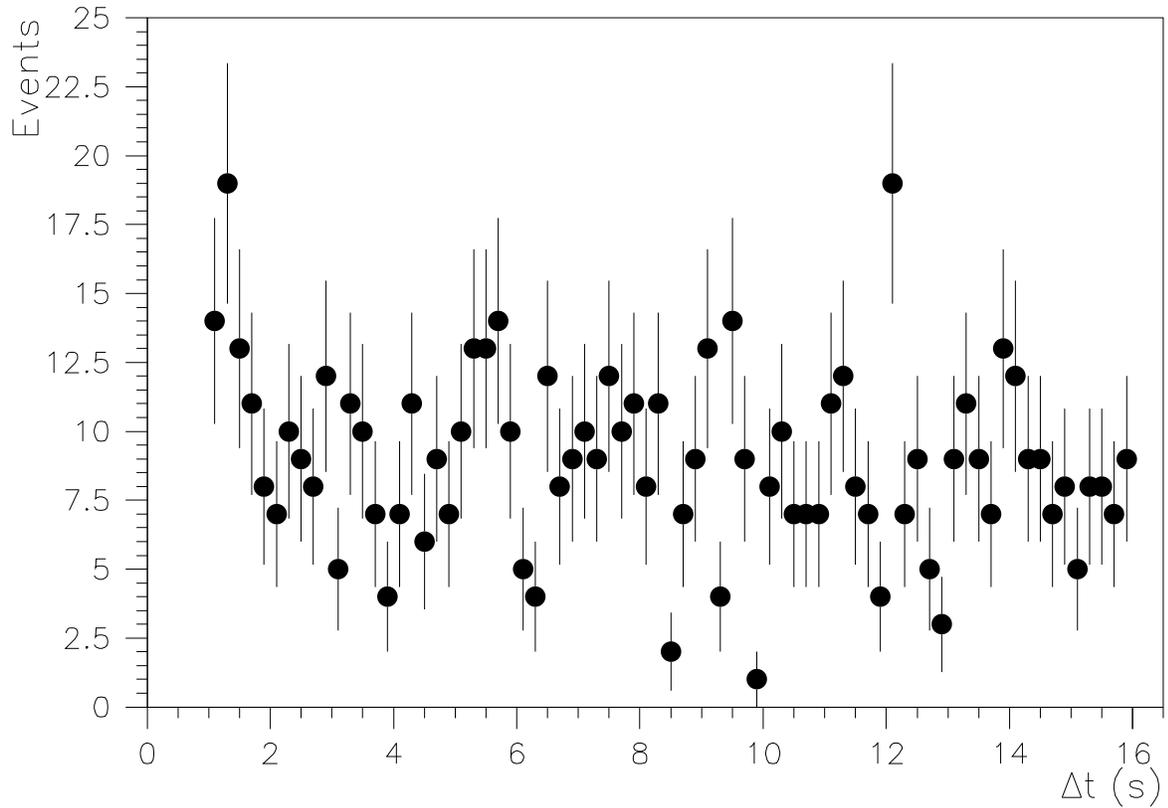,width=16cm}
}
\caption{Delay time ($\Delta$t) distribution for
candidate $\alpha$-$\alpha$ events for
$^{223}$Ra$\rightarrow$$^{219}$Rn$\rightarrow$$^{215}$Po
in DS$_5$. No time-correlated events are identified in
such that only an upper limit to the decay rates can be
set.
}
\label{dtu235}
\end{figure}

The sequence
DS$_{5a,5b}$ is another triple $\alpha$-cascade.
However, the second decay with
$\halflife$=1.78~ms is well inside
the DAQ dead time window (starting from
500~$\mu$s after the 
trigger for a duration of about 5~ms),
such that this cannot be measured.
The first decay, with
$\halflife$=3.96~s, is measurable.
The time-difference plot is shown
in Figure~\ref{dtu235}.
No evidence of an exponential decay
was observed, such that only an upper limit
to the $\ur235$ series can be derived.

\subsection{Radiopurity Levels }
\label{sec:radiopurity}

The measured intrinsic radiopurity levels
can be presented in several
ways and they
are summarized in Table~\ref{results}.
The measured activity in $\rm{mBq ~ kg^{-1}}$
can be converted to radiopurity
level in units of ``gram per gram (g/g)''
of the parent isotopes in CsI(Tl).
Assuming secular equilibrium,
the contamination levels of $\u238$
and $\th232$ can also be derived and
compared to other measurements. 
Results were derived from
the 1725~kg-day of data 
taken with 40 crystals,
and the quoted errors 
denote the uncertainties 
of the measured mean activities.

\begin{table}[hbt]
\begin{center}
\begin{tabular}{lcll}
\hline
DS & 
Measured Activity & Contaminations of & Contaminations of \\
& (mBq~kg$^{-1}$) 
& $~~$Long-Lived Parents (g/g )
& $~~$Series$^\dagger$ (g/g)\\ \hline \hline
1  & $0.0107 \pm 0.0004$ & 
$\rm{^{226}Ra :  2.92 \pm 0.11 \times10^{-19}}$ & 
$\rm{\u238 :  0.86 \pm 0.03 \times10^{-12}}$ \\
2  & $0.0102 \pm 0.0003$ & 
$\rm{^{226}Ra :  2.79\pm0.07\times10^{-19}}$ & 
$\rm{\u238 :  0.82\pm0.02\times10^{-12}}$ \\
3a,3b  &  $0.0090 \pm 0.0002$ & 
$\rm{^{228}Th :  2.97\pm0.08\times10^{-22}}$ & 
$\rm{\th232 : 2.23 \pm 0.06\times 10^{-12}}$ \\
4  &  $0.0061 \pm 0.0003$  &  
$\rm{^{228}Th : 3.1 \pm 0.2 \times 10^{-22}}$ &  
$\rm{\th232 : 2.3 \pm 0.1 \times 10^{-12} }$ \\
5a  &  $<0.003$ & 
$\rm{^{227}Th : < 1.6\times10^{-21}}$ & 
$\rm{\ur235 : < 4.9\times10^{-14}}$ \\ \hline
\multicolumn{4}{l}{$^\dagger$ assume secular equilibrium}
\end{tabular}
\caption{\label{results}
Measured activities and the derived
contaminations
of the four decay sequences. 
 }
\end{center}
\end{table}

\begin{figure} 
\centerline
{
\epsfig{figure=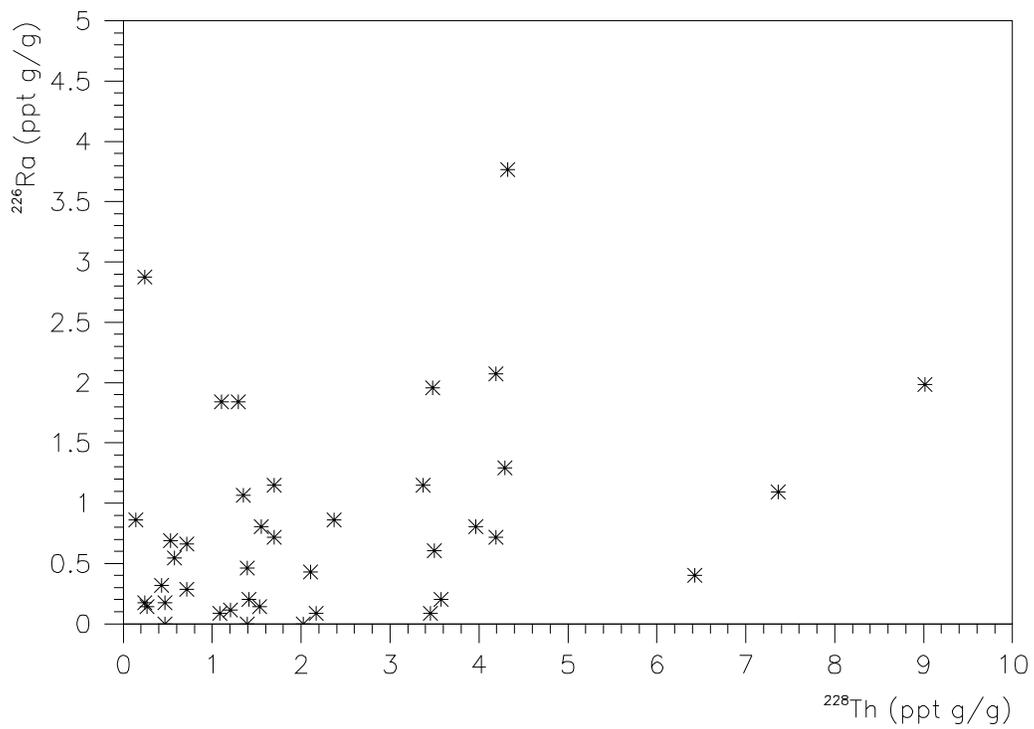,width=14cm}
}
\caption{
The correlations of 
$^{226}$Ra and $^{228}$Th 
contaminations among the
CsI(Tl) crystals.
}
\label{uVsth}
\end{figure}

The most accurate measurements were
derived from DS$_2$ and DS$_{3a,3b}$
giving, respectively, 
average activities of
$\rm{0.0102 \pm 0.0003 ~ mBq ~ kg^{-1}}$ and
$\rm{0.0090 \pm 0.0002 ~ mBq ~ kg^{-1}}$, corresponding
to contaminations of 
$\rm{2.79\pm0.07\times10^{-19} ~ g/g}$  and
$\rm{2.97\pm0.08\times10^{-22} ~ g/g}$  for
$^{226}$Ra and $^{228}$Th, respectively.
Assuming secular equilibrium,
the $\u238$ and $\th232$ concentrations
in CsI(Tl) are at the
$\rm{8.2 \pm 0.2 \times 10^{-13} ~ g/g}$  and
$\rm{2.23 \pm 0.06\times 10^{-12} ~ g/g}$
levels, respectively.
The scattered-plot for the $\u238$ and $\th232$
levels among the 40 crystals
are shown in Figure~\ref{uVsth}. 
The RMS spreads are, respectively,
80\% and 88\% for the
$^{226}$Ra and $^{228}$Th levels $-$
or alternatively the 
$\u238$ and $\th232$ contaminations
in the case of secular equilibrium. 
No strong correlations between the
two series were observed.

Comparison with other radiopurity measurements
in CsI(Tl) is given in Table~\ref{compare}.
The measured levels reported in this article
are consistent with and improve 
over independent measurements using
high purity germanium detectors on
samples of CsI powder.
Only upper limits of
$< 7.6 \times 10^{-9}$~g/g  and
$< 7.3 \times 10^{-10}$~g/g
could be derived
for the $\u238$ and $\th232$ series,
respectively,
from the absence of their
associated $\gamma$-lines.
The measurements are also consistent with the
previous results on CsI(Tl) crystals~\cite{csibkg,kims},
where the radiopurity levels were derived by
fitting the various $\alpha$-activities to
the measured $\alpha$-background spectra.
Reduction of the measurement uncertainties 
is due to the stringent selection 
criteria for
the time-correlated pairs which
greatly suppress the background.

\begin{table}[hbt]
\begin{center}
\begin{tabular}{lcccc}
\hline
Scintillators & Reference & Methods & $\u238$ & $\th232$ \\ 
& & & \multicolumn{2}{c}{( $10^{-12}$~g/g ) } \\ \hline \hline
CsI(Tl) & This Work & DS$_{2,3a,3b,4}$ & 0.82$\pm$0.02 & 2.23$\pm$0.06 \\
& This Work & HPGe & $<$ 7600 & $<$ 730 \\
& \cite{csibkg} & $\phi_\alpha$ & $<$6.0 & $<$19 \\ 
& \cite{kims} & $\phi_\alpha$ & 0.39$\pm$0.09 & 4.5$\pm$0.5 \\ \hline
NaI(Tl) & \cite{naidama} & DS$_2$ & 0.38 & 1.0 \\
 & \cite{naiuk} & DS$_{2,3a}$ & 4.0$\pm$0.2 & 2.4$\pm$0.5 \\
 & \cite{naijapan} & DS$_{2,3a}$ & 3.0$\pm$0.1  & 17$\pm$1  \\ \hline
CdWO$_4$ & \cite{cdwo4} & DS$_{2,3a}$ & $<$ 0.4  & 9.4$\pm$0.7  \\ \hline
GSO(Ce) & \cite{gso} & DS$_{2,3a}$ & 110$\pm$20 
& 2.9$\pm$0.3$\times 10^{4}$ \\  \hline
CeF$_3$ & \cite{cef3} & DS$_{2,3a}$ & $< 5.6 \times10^3$ 
& $1.4 \times 10^4$ \\ \hline
BaF$_2$ & \cite{baf2} & DS$_{2,4}$ & $1.1\times10^5$ & $10^5$ \\ \hline
CaF$_2$(Eu) &  \cite{caf2} & DS$_{2,3a}$ & 
8920$\pm$80 & 24.3$\pm$0.4 \\ \hline \hline
Liquid & & & & \\
Scintillator & \cite{ctf} & DS$_{2,3a}$ 
& $3.2 \times 10^{-4}$ & $< 2.5 \times 10^{-4}$ \\  \hline
\end{tabular}
\caption{ \label{compare}
A summary of the measurements of $\u238$ and $\th232$ concentrations
in various crystal scintillators,
assuming secular equilibrium within the series in all cases.
Most results are due to measurements of the
correlated events from the various
decay sequences.
HPGe denotes measurements 
of $\gamma$-peaks with a high-purity
germanium detector, while $\phi_{\alpha}$ 
represents statistical fits of the
purity levels with 
the measured $\alpha$-spectra.
}
\end{center}
\end{table}

A comparison of the measured intrinsic radiopurity
levels of $\u238$ and $\th232$ with those
in other crystal scintillators as well
as in the kilo-ton low-background liquid scintillator
CTF (Counting Test Facility)
detector~\cite{ctf} at Gran Sasso is given in
Table~\ref{compare}.
The purity of the CsI(Tl) crystals
with respect to these series
is comparable with
the best measured for NaI(Tl) and CdWO$_4$ and
exceeds those of the other crystals.
Nevertheless, contaminations in all crystals
are much higher than those in liquid
scintillator.

The sequences (DS$_1$, DS$_2$)
and (DS$_{3a,3b}$, DS$_4$) originate from the
same long-lived parent isotope: $^{226}$Ra
and $^{228}$Th, respectively. 
As shown in Table~\ref{results}
The agreement between their measured activities 
are excellent.
Alternatively,
the ratios between DS$_4$ to DS$_{3a,3b}$ 
give the branching ratio of the
$\rm{ ^{212}Bi \rightarrow ^{212}Po \rightarrow ^{208}Pb}$
branch within the $\th232$ series.
The measured branching ratio is 68$\pm$4\%,
consistent with the expected value of 64\%.

The absence of signals for DS$_{5a}$, as shown in
Figure~\ref{dtu235}, was converted into
upper limits of $\rm{0.003 ~ mBq ~ kg^{-1}}$
in the measured activity, corresponding to
$1.6\times10^{-21}$~g/g
in the contaminations of $^{227}$Th.
Equilibrium within $\ur235$ series 
imply an upper limit of
$4.9\times10^{-14}$~g/g can be derived for
$\ur235$ contaminations in the CsI(Tl)
crystals.

Given in Table~\ref{sensitivity} is
an alternative view to assess 
the merits of our measurements by
{\it projected} sensitivities 
in the hypothetical scenario where no
correlated events are observed and
only upper limits can be derived.
This was indeed the case for DS$_{5a}$.
The accidental rates are due to uncorrelated
event-pairs  which survive the various
selection cuts and fall into 
time windows corresponding 
to four half-lives.
One can derive from the rates
the limiting sensitivities in
$\rm{mBq ~ kg^{-1}}$, as well
as the upper limits of the radiopurity
levels of the
decay parents and the two series.
The limits represent the levels above
which the $\u238$/$\th232$ contaminations 
of a test sample can be detected by 
the present methods
at the measured accidental background rates.
The restrictive selection criteria 
adopted in this work 
(ID+$\Delta$t+$\Delta$Z+E$_{\alpha}$)
greatly suppress the accidental
background and enhance the 
the detection sensitivities 
for the correlated decay sequences
in DS$_{2,3a,3b,4}$.
The achieved levels of $\sim 10^{-16}$~g/g 
are comparable to those of massive
liquid scintillator~\cite{ctf} listed
in Table~\ref{compare}.

\begin{table}[hbt]
\begin{center}
\begin{tabular}{lccll}
\hline
DS & Accidental Rate &
Limiting Sensitivity & Contaminations of & Contaminations of \\
& (kg$^{-1}$day$^{-1}$) & (mBq~kg$^{-1}$) 
& $~~$Long-Lived
& $~~$Series$^\dagger$ (g/g)\\ 
& & &  $~~$Parents (g/g ) & \\ \hline \hline
1 & 0.53  &  $6.2 \times 10^{-3}$ & 
$\rm{^{226}Ra :  < 1.7 \times 10^{-19}} $ & 
$\rm{\u238 :  < 5.1 \times 10^{-13} }$ \\
2 & 0.0029 &  $3.3 \times 10^{-5}$ & 
$\rm{^{226}Ra :  < 9.0 \times 10^{-22}} $ & 
$\rm{\u238 :  < 2.7 \times 10^{-15} }$ \\ \hline
\multicolumn{4}{l}{ ~~~ Combined for $\u238$ Series:} &  
$\rm{\u238 :  < 2.7 \times 10^{-15} }$ \\ \hline \hline
3a,3b & $7.5 \times 10^{-5}$ & $8.5 \times 10^{-7}$  & 
$\rm{^{228}Th :  < 2.9 \times 10^{-26} }$ & 
$\rm{\th232 : < 2.2 \times 10^{-16}}$ \\
4 & $1.6 \times 10^{-5}$ &   $1.9 \times 10^{-7}$ &  
$\rm{^{228}Th :  < 9.4 \times 10^{-27} }$ & 
$\rm{\th232 : < 7.0 \times 10^{-17}}$ \\ \hline
\multicolumn{4}{l}{ ~~~ Combined for $\th232$ Series:} &  
$\rm{\th232 :  < 5.3 \times 10^{-17} }$ \\ \hline \hline
5a & 0.26 &  0.003 & 
$\rm{^{227}Th : < 1.6 \times 10^{-21}}$ & 
$\rm{\ur235 : < 4.7 \times 10^{-14}}$ \\ \hline
\multicolumn{4}{l}{$^\dagger$ assume secular equilibrium}
\end{tabular}
\caption{\label{sensitivity}
Projected limiting sensitivities and the contamination
levels of the five decay sequences in the case where
no correlated signals are observed.
}
\end{center}
\end{table}

These cascade events were used to provide
measurements of several important parameters
on crystal properties and detector performance.
In the following sub-sections, 
we adopted for analysis a subset of
the crystals 
(6 and 13 crystals out of 40 
for the $\u238$ and $\th232$ series,respectively)
which have sufficiently
large event samples to provide statistically
meaningful measurements.

\subsection{Quenching Factor Measurements}
\label{subsec::qf}

The ionization charge densities are much higher 
for $\alpha$-particles than
those for electrons or photons, giving
rise to enhanced recombination rates and therefore
the light yield is reduced.
The quenching factor(QF) of a scintillator is
the ratio of measurable light yield 
compared to  that due to 
electrons/photons of
the same energy:
\begin{equation}
\rm{
QF ~ = ~ \frac{E_{ee}}{E_{Q}}
}
\end{equation}
where $\rm{E_{ee}}$ is  the
measurable ``electron-equivalent'' energy
of the $\alpha$-particles and 
$\rm{E_{Q}}$ is Q-value of decays which
is also the kinetic energy of the $\alpha$'s.
The quenching is energy-dependent, and
is usually described by two parameters (a,b)
in the form~\cite{scinbasic}:
\begin{equation}
\rm{
E_{ee} ~ = ~ \frac{a \cdot  E_Q}{1 + b \cdot  E_{Q}} ~~.
}
\end{equation}

The identified $\alpha$-events from the correlated
pairs provide a background-free
sample for the measurement
of QF. The value of $\rm{E_{Q}}$ is exactly 
known and the entire
volume of the crystal is sampled, 
as compared to measurements
with an external $\alpha$-source which 
are sensitive only to a 
localized sites on the crystal surface.

Tabulated in
Table~\ref{qf} are the measured QF for 
the five $\alpha$'s from the various
cascades studied in this work.
The $\alpha$'s from DS$_1$,
DS$_2$,  DS$_{3a,3b}$ 
are isolated events and their light yield
can be readily measured by integrating
the pulses. The energy measurements for
$\alpha$'s in DS$_4$, however, involve
de-convolutions of the $\beta$-$\alpha$ pulses
shown in Figure~\ref{ds4}a, and therefore
are subjected to bigger uncertainties.
The energy resolutions of the $\alpha$-events
($\rm{\sigma ( E_{ee} ) / E_{ee} }$) are also shown,
where $\rm{\sigma ( E_{ee} )}$ is the 
RMS of the
respective peaks shown in Figure~\ref{alphaspectrum}.

\begin{table}[hbt]
\begin{center}
\begin{tabular}{cccccc}
\hline
DS & Decays & $\rm{E_Q}$ (MeV) &  
$\rm{E_{ee}}$ (MeV) & QF & 
$\rm{\sigma (E_{ee}) / E_{ee}}$ \\ \hline \hline
1 & $^{222}$Rn $\rightarrow$ $^{218}$Po  
&  5.590  &  2.76$\pm$0.02    &  0.494$\pm$0.004 &  0.076 \\
3a & $^{224}$Ra $\rightarrow$ $^{220}$Rn  
&  5.789  &  2.92$\pm$0.02    &  0.504$\pm$0.003 &  0.074 \\
1 & $^{218}$Po $\rightarrow$ $^{204}$Pb  
&  6.115 &  3.09$\pm$0.02    &  0.505$\pm$0.003 &  0.097 \\
3a & $^{220}$Rn $\rightarrow$ $^{216}$Po  
&  6.405  &  3.33$\pm$0.02    &  0.520$\pm$0.003 & 0.069 \\
3b & $^{216}$Po $\rightarrow$ $^{212}$Pb  
&  6.907  &  3.67$\pm$0.02    &  0.532$\pm$0.003 & 0.068 \\
2 & $^{214}$Po $\rightarrow$ $^{210}$Pb  
&  7.834  &  4.33$\pm$0.03    &  0.553$\pm$0.003 & 0.11 \\
4 & $^{212}$Po $\rightarrow$ $^{208}$Pb  
&  8.954  &  5.08$\pm$0.05    &  0.568$\pm$0.006 & 0.14 \\ \hline
\end{tabular}
\caption{\label{qf}
The measured quenching factors due to 
$\alpha$-emissions in the various decay
sequences.}
\end{center}
\end{table}

It can be seen that 
the quenching behaviour are different
for  inorganic crystal scintillators
and organic liquid scintillators.
The QF is large in crystal scintillators
($\sim 0.5$ for CsI(Tl) for $\alpha$-particles in the
MeV range)
compared to that for liquid scintillator,
typically at the QF$<$0.1  range~\cite{qfliqscin}.
In addition, the quenched pulses for $\alpha$-particles
have faster decay times compared to those from electrons/photons
in crystal scintillators, as indicated in Figure~\ref{psd},
while the reverse is true for liquid scintillators.
The quenching for nuclear recoils in CsI(Tl) at the 
keV energy range is relevant to Dark Matter searches, 
and have been measured (QF$\sim$0.2) 
by various groups~\cite{kims,qfrecoil}.

\begin{figure}
{\bf (a)}\\
\centerline{
\epsfig{figure=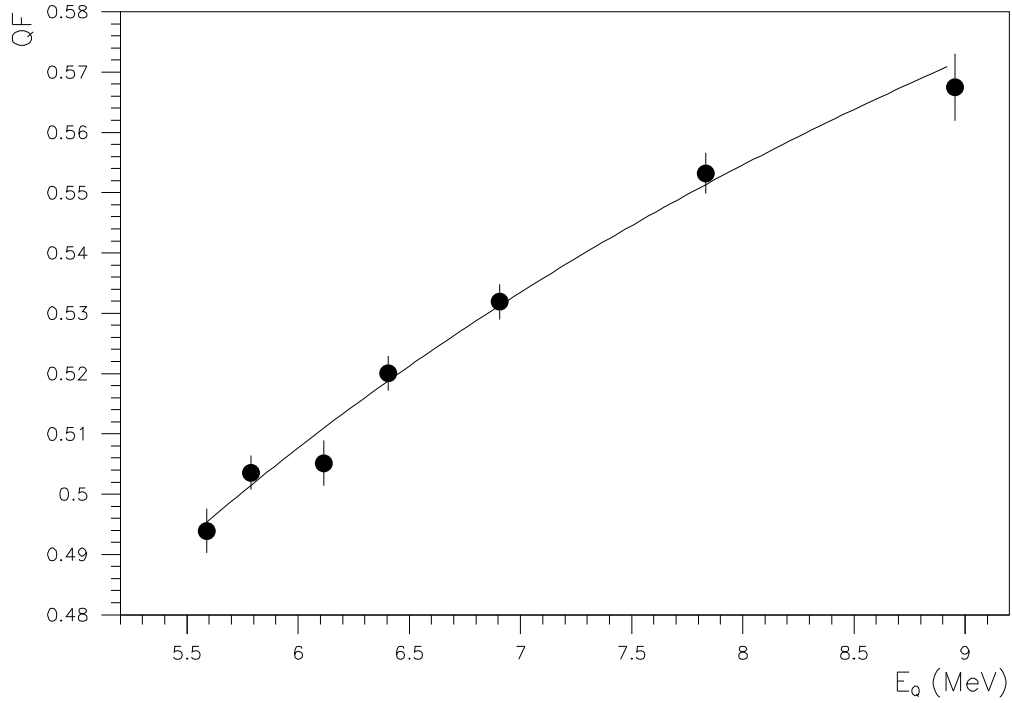,width=14cm}
}
{\bf (b)}\\
\centerline{
\epsfig{figure=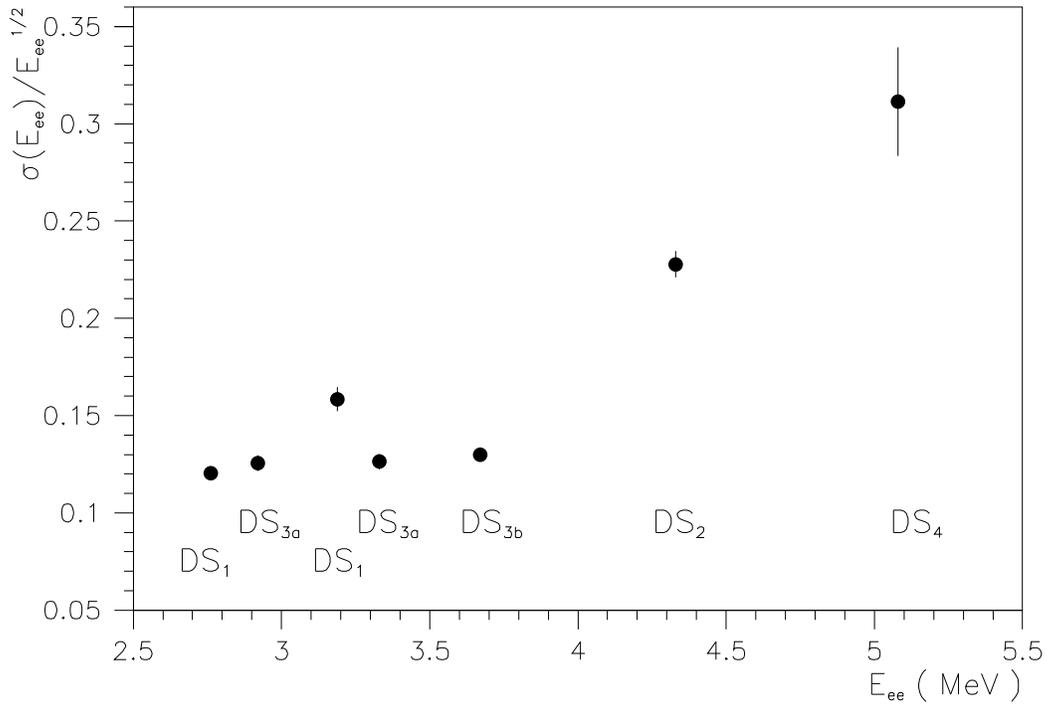,width=14cm}
}
\caption{The variations of (a) the measured quenching factors
in CsI(Tl)
with the kinetic energy of the $\alpha$-particles,
and (b) the figure-of-merit
$\rm{\sigma ( E_{ee} ) / \sqrt{ E_{ee} } }$ 
with the measured energy $\rm{ E_{ee} }$.
See text for explanations of the worse resolution
in three of the data points.
}
\label{fig:qf}
\end{figure}

The variations of QF with the
kinetic energy  of the $\alpha$'s
are displayed
in Figure~\ref{fig:qf}a.
The measured quenching parameters
a=0.25$\pm$0.01 and b=0.33$\pm$0.02~MeV$^{-1}$
provide good fits to the data.
The RMS variation for
the QF among the 13~crystals studied
is at the 6.3\% level.
The variations  of 
$\rm{ \frac{\sigma (E_{ee})}{\sqrt{E_{ee}} }}$ with 
$\rm{E_{ee}}$ are
displayed in Figure~\ref{fig:qf}b.
The energy resolution is 
statistically-limited in four
of the data points where
$\rm{\sigma (E_{ee})} \propto \sqrt{E_{ee}}$. 
The worse resolution
of the $\alpha$ from $^{218}$Po in DS$_1$ 
is due to its long half-life such that
the sample is contaminated by accidental
events.
Energy measurement of the 
$^{214}$Po-$\alpha$ in DS$_2$ is affected by
the sampling procedures of this
particular data acquisition system 
$-$ the pedestals for the delayed events
within 500~$\mu$s of the primary trigger
were not recorded and uncertainties
were therefore introduced.
Measurement of the $^{212}$Po-$\alpha$ 
in DS$_4$ involves
the de-convolution of the $\beta$-$\alpha$ pulse
shapes, which gives additional contributions
to the energy resolution.

\subsection{Spatial Resolution}
\label{subsec:stcorrelation}

\begin{figure}
\centerline{
\epsfig{figure=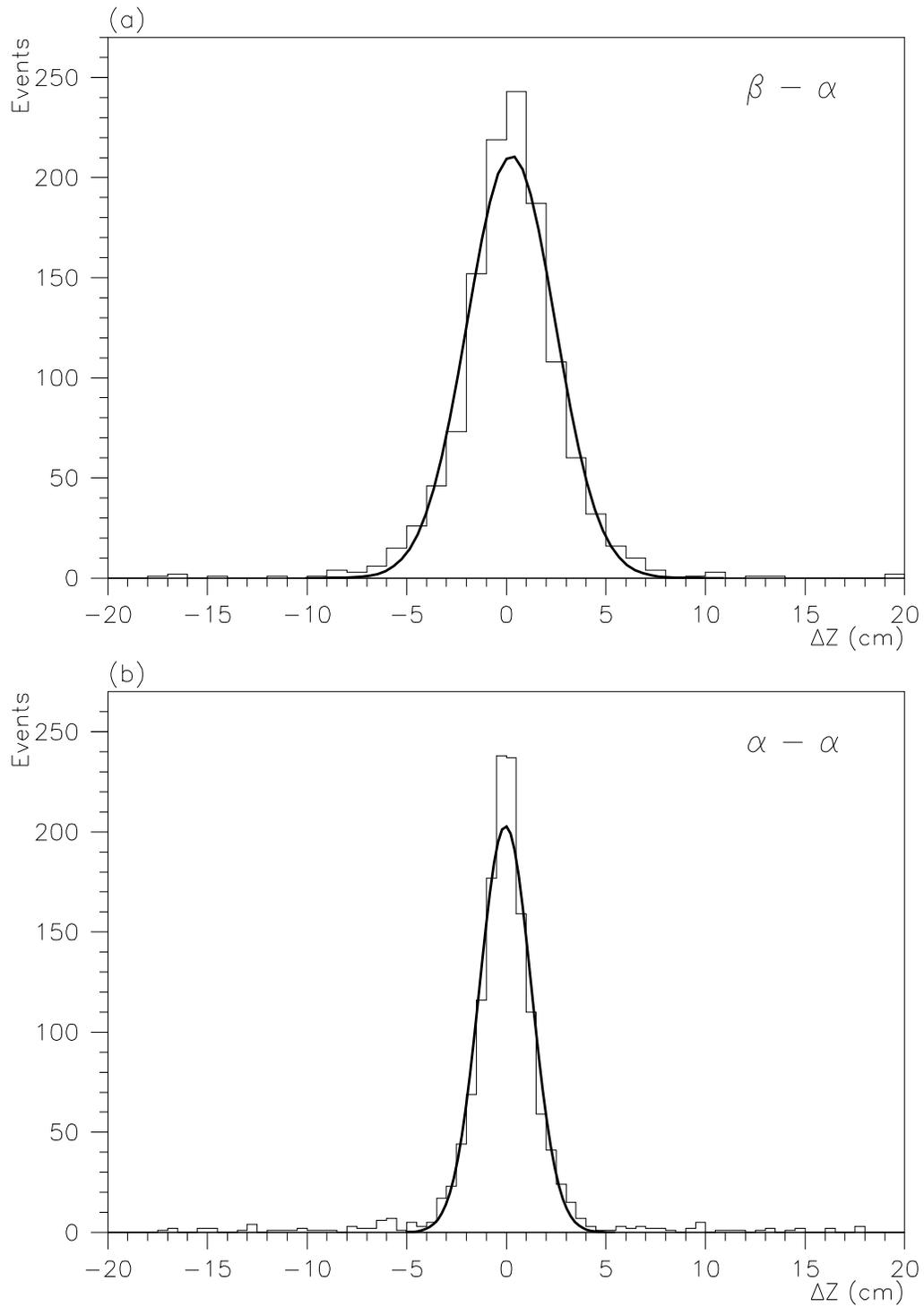,width=14cm}
}
\caption
{The $\Delta$Z-distributions 
from a typical CsI(Tl) crystal on
(a) $\beta$-$\alpha$
and
(b) $\alpha$-$\alpha$ events, from which 
the Z-resolutions $\sigma _Z$ can be derived.
}
\label{deltaz}
\end{figure}

The longitudinal ``Z-position'' of an event
can be derived from the difference of the
PMT signals between the {\it left} ($\rm{Q_L}$)
and {\it right} ($\rm{Q_R}$) sides~\cite{proto,erecon}:
\begin{equation}
\rm{
R ~ = ~  \frac{1}{2} ( \frac{ Q_L - Q_R }{Q_L + Q_R} )
} \end{equation} 
where R is a dimensionless parameter linearly
proportional to Z.
Data taken with
external collimated $\gamma$-sources prior to
installation provided measurements of
the Z-resolutions ($\sigma_Z$=2~cm) above 400~keV.
This approach is limited by the intrinsic
spread of the photon interaction sites 
within the crystal due to multiple Compton
scatterings.
A better in situ method is through
the study of the measured position
difference ($\Delta$Z) of 
the correlated $\beta$-$\alpha$ and
$\alpha$-$\alpha$ pairs which are emitted
at the same site.
The $\Delta$Z distributions of $\beta$-$\alpha$
from DS$_2$ and the two $\alpha$-$\alpha$
pairs 
from DS$_{3a,3b}$ for the combined data
are depicted in Figure~\ref{deltaz}a\&b.
Both are centered at $\Delta$Z=0, indicating that
the selected pair of events were indeed
originated at the same site.
The resolution $\sigma_Z$, as 
given by the RMS of the distributions,
are 2.2~cm 
and 1.3~cm for the $\beta$-$\alpha$ 
and $\alpha$-$\alpha$ events, respectively,
while
the RMS spread of $\sigma_Z$ among
the crystal samples are
19\% and 24\%. 

The $\alpha$-$\alpha$ samples provide a
more accurate description in the studies
of the intrinsic
spatial resolution, since both of the
$\alpha$-pairs are originated at the same
site. The $\beta$-events
in DS$_2$, on the other hand, 
are accompanied
by other $\gamma$-emissions 
such that their exact vertices
are not well-defined.
This explains the worse $\Delta$Z
distribution for the  $\beta$-$\alpha$
samples.

\subsection{Contamination Gradient with Crystal Growth Axis}
\label{subsec:crygrowth}

\begin{figure}
\centerline{
\epsfig{figure=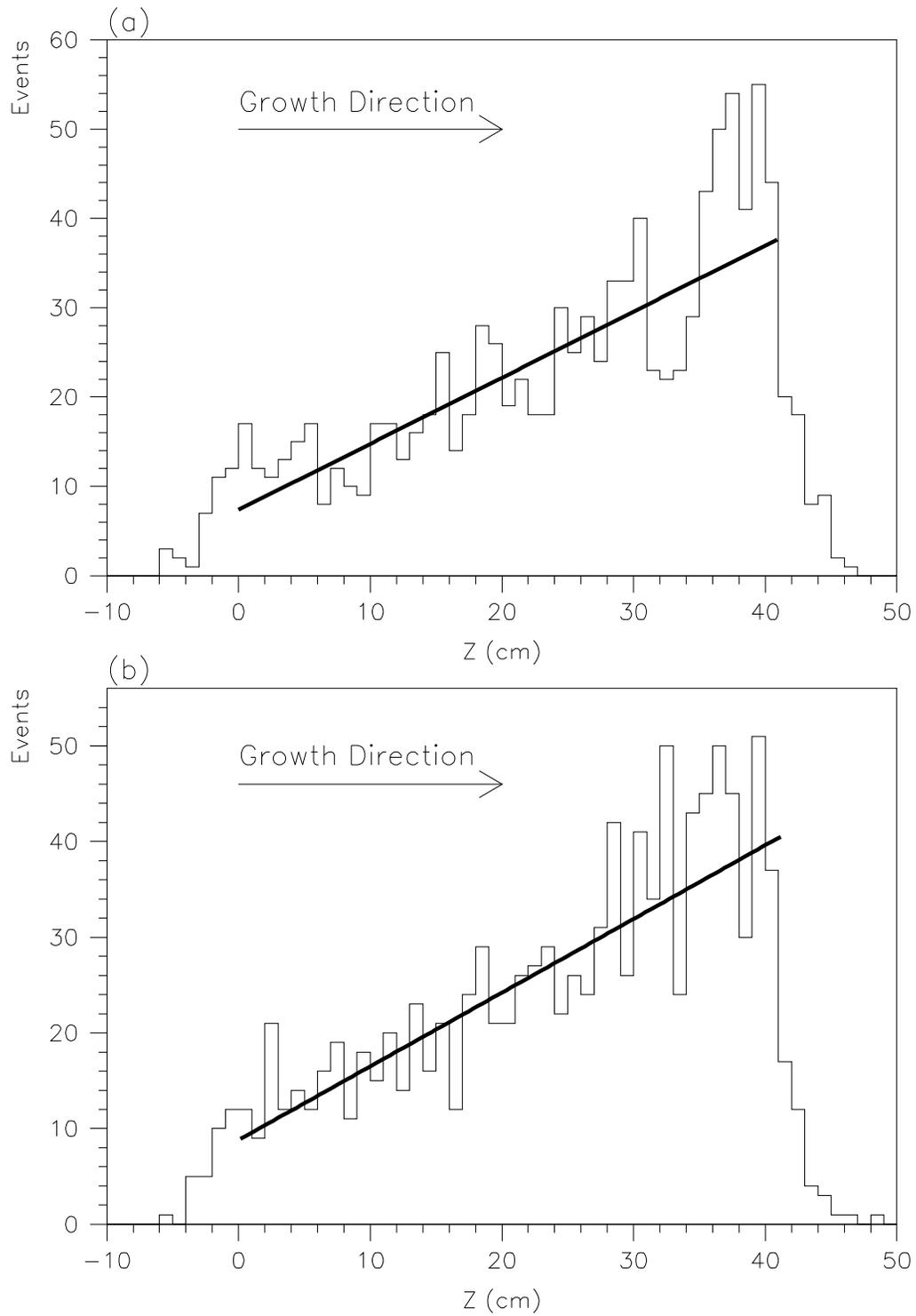,width=14cm}
}
\caption
{The typical Z-position distributions of the radiopurity 
contaminations from 
(a) $^{226}$Ra within the $\u238$ series 
and 
(b) $^{228}$Th within the $\th232$ series.
}
\label{gradient}
\end{figure}

Crystal growth is itself an
purification process.
As a result, the impurity concentrations 
in the solution
increase during the growth process,
such that residual 
contaminations inside the crystals
are expected to increase with the crystal 
growth axis.
This is verified in 
Figure~\ref{gradient}a\&b 
where the Z-position distributions of 
the event rates of DS$_2$ and DS$_{3a,3b}$
are depicted,
corresponding respectively to the
gradients of the $^{226}$Ra and $^{228}$Th
levels.
The measured gradients of the combined
data are
$\rm{ 6.8\times10^{-21} ~g/g \cdot cm}$
for  $^{226}$Ra and
$\rm{ 1.6\times10^{-23} ~ g/g \cdot cm}$
for $^{228}$Th, while the
RMS spread among the samples are
33\% and 104\%, respectively.
This behaviour is in contrast to the
uniform Z-position distribution
of $^{137}$Cs
shown in Figure~\ref{cs137}b,
since the trace contaminations
in $^{137}$Cs
cannot be separated from the
natural $^{133}$Cs by any chemical
means.

\subsection{Data Acquisition Dead Time}

The event samples can also be used to
measure or monitor the data acquisition
dead time (DT).
The fraction DT/(1-DT) is the
ratio of recorded double-$\alpha$ events 
in DS$_{3a}$ where the third one is missing to
the triple-$\alpha$ events where the
entire sequence DS$_{3a}$ and
DS$_{3b}$ are reconstructed.
The measured value is DT=9$\pm$1\%,
in good agreement to DT=9.5\%  
derived by 
an alternative method using random trigger
events~\cite{eledaq}.

\section{Summary and Conclusion }

We report in this article measurements
of the intrinsic radiopurity in CsI(Tl).
In addition to energy measurements 
and particle identifications,
the selection of spatially
and temporally correlated event-pairs
within the $\u238$ and $\th232$ series
can  greatly suppress
accidental background and enhance the
measurement capabilities.
Sensitivity levels comparable 
to dedicated massive
low-background liquid scintillator detector
were achieved.
In particular, the $\alpha$-pairs 
correlated by time as long as 10~minutes
were successfully identified.
The CsI(Tl) crystals measured in this work
are among the cleanest compared to the other 
crystal scintillators in the $\u238$
and $\th232$ contaminations.
The methods and results 
will be valuable references 
to other low-background
experiments where the suppression
and measurements of the $\u238$
and $\th232$ background are necessary.

We are grateful to the referee for pointing
out a numerical error.
This work was supported by contracts
92-2112-M-001-057 and
93-2112-M-001-030
from the National Science Council, Taiwan,
and 19975050 from the
National Science Foundation, China.


\newpage

\end{document}